# Emergence of giant spin-orbit torque in a two-dimensional hole gas on the hydrogen-terminated diamond surface

Fujio Sako[1,#], Ryo Ohshima[1,2,#], Yuichiro Ando[1-3], Naoya Morioka[2,4], Hiroyuki Kawashima[4], Riku Kawase[4], Norikazu Mizuochi[2,4], Hans Huebl[5-7], and Masashi Shiraishi[1,2]

1. Department of Electronic Science and Engineering, Kyoto University, Kyoto, Kyoto 615-8510, Japan.

2. Center for Spintronics Research Network (CSRN), Institute for Chemical Research, Kyoto University, Uji, Kyoto 611-0011, Japan.

3. PRESTO, Japan Science and Technology Agency, Honcho, Kawaguchi, Saitama 332-0012, Japan

4. Institute for Chemical Research, Kyoto University, Uji, Kyoto 611-0011, Japan.

5. Walther-Meißner-Institut, Bayerische Akademie der Wissenschaften, 85748 Garching, Germany.

6. TUM School of Natural Sciences, Technical University of Munich, 85748 Garching, Germany.

7. Munich Center for Quantum Science and Technology (MCQST), 80799 Munich, Germany.

# These two authors contributed equally to this work.

Corresponding author:

Ryo Ohshima (ohshima.ryo.2x@kyoto-u.ac.jp)

Masashi Shiraishi (shiraishi.masashi.4w@kyoto-u.ac.jp)

**Abstract**

Two-dimensional (2D) carrier systems exhibit various significant physical phenomena for electronics and spintronics, where one of the most promising traits is efficient spin-to-charge conversion stemming from their Rashba-type spin-orbit interaction. Meanwhile, a nuisance in quests of promising materials for spintronics application is that vast majority of the investigated platforms consists of rare and/or toxic elements, such as Pt and Te, which hinders progress of spin conversion physics in view of element strategy and green technology. Here, we show the emergence of giant spin-orbit torque driven by 2D hole gas at the surface of hydrogen-terminated diamond, where the constituent substances are ubiquitous elements, carbon and hydrogen. The index of its spin torque efficiency at room temperature is several times greater than that of rare metal, Pt, the benchmark system/element for spin-to-charge conversion. Our finding opens a new pathway for more sustainable spintronics and spin-orbitronics applications, with efficient spin-orbit torque employing ubiquitous non-toxic elements.

Two-dimensional (2D) carrier systems hosting abundant intriguing condensed-matter physics have been collecting tremendous attention and been intensively investigated. In the advent of the systems, III-V compound semiconductor systems, such as the AlGaAs/GaAs heterostructure[1], were vigorously studied, and a high mobility two-dimensional electron gas (2DEG) created in the system enabled fabrication of high frequency electron devices. The discovery of the $LaAlO_3$/$SrTiO_3$ (LAO/STO) oxide-based 2DEG opened a new era of 2D electron systems, because a high mobility electron layer consisting of *d*-electrons can be created in between oxide insulators[2], which challenged conventional approaches to create 2DEGs using semiconductors heterostructures. Notably, the 2DEG at LAO/STO interfaces can host a wide variety of attractive physical traits, such as ferromagnetism, superconductivity and even their coexistence[3–8], tuneable spin-to-charge conversion[9], and room temperature spin propagation[10] rendering this system highly attractive for the investigation of 2DEGs. In addition, these pivotal properties motivated investigation of derived 2DEG system by replacing LAO with $LaTiO_3$ which resulted in the demonstration of a giant spin-to-charge conversion mechanism originating from the Rashba-Edelstein length of 190 nm at 15 K, however the effect was absent over 100 K[11]. Notably, most two-dimensional charge carrier systems are localized at interfaces between two materials, which enables the required spatial confinement. While beneficial for applications like transistors, the typical insulating properties of the top layer present an obstacle when the galvanic coupling to the two-dimensional charge carrier system is desired.

In the history of 2D carrier systems, the vast majority of studies focused on electron-based systems, i.e., 2DEG, and indeed, only a limited number of studies focused on the creation of two-dimensional hole gases (2DHG)[13–16], the complementing two-dimensional carrier system. Hydrogen-terminated diamond (H-diamond) has the unique property to host a 2DHG on its surface even without doping[13,17–19] (see Fig. 1(a)), which allows direct electrical access to the 2DHG system and circumventing the problem holding most of the 2D carrier systems described above. Considering the possible existence of a strong Rashba field[20–22], investigating novel spintronics functions of the 2DHG in H-diamond that remained elusive so far is quite significant. The estimated spin-splitting energy $\Delta_{SO}$ in H-diamond is reported to be several meV[20], which even exceeds that of other 2DEG systems such as strained InGaAs and Ge quantum wells[23–25] and is on-par with that of LAO/STO interfaces[26]. Therefore, the presence of the Rashba-Edelstein effect can be contemplated, which is induced by the sizable Rashba SOI (see Fig. 1(b)). The latter is also linked to the quite efficient spin conversion, which is initially surprising as diamond is a carbon-based material with high spatial inversion symmetry. Furthermore, the conventional strategy to realize efficient spin conversion resorts to utilization of rare and/or toxic elements, such as Pt, Sr (for oxide 2D carriers), Bi Se, and Te (for topological materials), which is a huge nuisance for establishing novel green technologies in spintronics and spin-orbitronics.

In this work, we delve into the subject by focusing on the spin-orbit torque (SOT) of the 2DHG at the H-diamond surface. Giant damping-like torque was successfully observed in a H-diamond/$Ni_{80}Fe_{20}$ (Py) system at room temperature (RT), and in fact, charge-spin conversion efficiency for the Rashba-Edelstein effect, Rashba-Edelstein effect length in H-diamond is 0.19 nm, which is greater than that of heavy metals such as Pt[27] and is a surprisingly high efficiency given that H-diamond constitutes of merely light elements like carbon and hydrogen. The results allow envisaging a wide variety of environmental-friendly spintronics applications using ubiquitous materials, H-diamond, towards such as SOT-MRAM without utilizing rare, expensive and toxic elements such as Pt, Bi, Se, and so forth.

A chemical-vapor-deposition (CVD)-grown undoped type-IIa (001) diamond (Augsburg Diamond Technology GmbH) was exposed to hydrogen plasma at 700°C, 20 kPa for 10 min. following surface polishing by diamond powder. The 2DHG was formed by using the diamond that was exposed to hydrogen (see also Methods and Fig. 1(c)). The surface roughness after the hydrogen-termination was estimated to be 0.15 nm in root-means square by atomic-force microscopy[28]. A Hall-bar structure was formed by making a mask by electron-beam (EB) lithography and reactive ion etching with oxygen plasma for 15 sec. in 50 W in plasma power. We confirmed that the surface of diamond after exposing to oxygen plasma was insulating due to the oxygen-termination of the surface[28]. The sheet conductivity and sheet carrier density of the H-diamond were estimated to be $2.8 \times 10^{-2}$ mS/□ and $3.2 \times 10^{12}$ cm$^{-2}$ by the current-voltage characteristics and Hall measurement, respectively. The estimated values are comparable to the other previous studies[29,30], which is evidence that the 2DHG is successfully formed at the surface of the H-diamond. The Py films with MgO(2 nm)/AlO$_x$(2 nm) capping layers were fabricated on the H- and O-diamonds by using EB lithography and EB deposition (see Fig. 1(c)). The capping layers were deposited *in situ* after the Py deposition, and the top Al layer was naturally oxidized. The harmonic Hall measurement was carried out by injecting an ac current with an amplitude of 1 mA for devices of $t_{Py} < 4$ nm and 3 mA for the other devices by using the AC source meter (Keithley 6221). The frequency was set to be 17 Hz. An external magnetic field $H_{ext}$ was applied along the out of plane direction of the 2DHG for the AHE measurement and applied along the in-plane direction for the PHE and the second harmonic Hall measurements by using Physical Property Measurement System (Quantum Design). Hall voltages were measured by using the lock-in amplifier (Stanford Research Systems SR830). All measurements were carried out at room temperature.

To detect the spin torque and estimate its amplitude, we exploited the second harmonic measurement, the anomalous Hall effect (AHE), and planar Hall effect (PHE) measurements for the 4 nm-thick Py device. This allows us to estimate the spin torque amplitude quantitatively. The amplitude of the AHE $V_{AHE}$ and the PHE $V_{PHE}$ were estimated to be −3.35 mV and 0.77 mV, respectively, and the out-of-plane anisotropy field $H_K$ was estimated to be $\mu_0 H_K = 0.46$ T, where $\mu_0$ is the vacuum permeability[28]. Figure 2(a) shows the azimuth-angular $\phi$ dependence of the second harmonic Hall voltage $V^{2\omega}$, where the result was fitted using the following equation[31–33]:

$$V^{2\omega} = A(H_{\text{ext}}) \cos\phi + B(H_{\text{ext}}) \cos 2\phi \cos\phi$$

$$= \frac{1}{2}\left(V_{\text{AHE}} \frac{H_{\text{DL}}}{H_{\text{K}}+H_{\text{ext}}} + V_{\text{ANE}} + V_{\text{ONE}} H_{\text{ext}}\right)\cos\phi + V_{\text{PHE}}\frac{H_{\text{FL}}+H_{\text{Oe}}}{H_{\text{ext}}}\cos 2\phi \cos\phi, \quad (1)$$

where $H_{\text{DL}}$ and $H_{\text{FL}}$ are the damping-like torque and field-like torque fields, respectively, $H_{\text{Oe}}$ is the Oersted field, $V_{\text{ANE}}$ and $V_{\text{ONE}}$ are the amplitude of the anomalous Nernst effect and the ordinary Nernst effect, respectively. As can be seen, the experimental result is nicely reproduced by the conventional fitting function. Figure 2(b) shows the $\mu_0 H_{\text{ext}}$ dependence of $A(H_{\text{ext}})$ in Eq. (1). $A(H_{\text{ext}})$ was deconvoluted into $V_{\text{AHE}}$, $V_{\text{ANE}}$, and $V_{\text{ONE}}$ components by Eq. (1) and $H_{\text{DL}}$ was estimated to be $\mu_0 H_{\text{DL}} = -97$ μT. The discernible damping-like torque field unequivocally indicates successful creation of the SOT in the H-diamond/Py system. The $\mu_0 H_{\text{ext}}$ dependence of $B(H_{\text{ext}})$ in Fig. 2(c) enables us to estimate $\mu_0(H_{\text{FL}} + H_{\text{Oe}}) = -13$ μT. Albeit $\mu_0(H_{\text{FL}} + H_{\text{Oe}})$ is substantially small, the oppositely aligned $H_{\text{FL}}$ and $H_{\text{Oe}}$ that are determined by considering the Rashba field generated in the H-diamond/Py interface as shown in Fig. 1(b) gives rise to the small magnitude of $\mu_0(H_{\text{FL}} + H_{\text{Oe}})$. Since the damping-like toque can provide dominant contribution to magnetization switching in SOT devices, the $H_{\text{DL}}$ is investigated in detail in the following sections.

To estimate the spin torque efficiency quantitatively, the harmonic Hall measurements were carried out for various thickness $t_{\text{Py}}$ of the top Py layer on hydrogen and oxygen terminated diamond (H- and O-diamond, respectively), which host a 2DHG and show insulating properties, respectively. Figure 3(a) shows the $H_{\text{ext}}$ dependence of $A(H_{\text{ext}})$ in the O-diamond/Py(5 nm) device, where the $A(H_{\text{ext}})$ exhibits a linear dependence to the $H_{\text{ext}}$, resulting in the negligibly small $H_{\text{DL}}$. The absence of the spin torque in O-diamond is unequivocally ascribed to insulating properties of O-diamond. Figures 3(b) and 3(c) show the $H_{\text{ext}}$ dependence of the $A(H_{\text{ext}})$ observed for H-diamond with the Py(5 nm) and Py(7 nm), respectively. The $A(H_{\text{ext}})$ in the H-diamond/Py stack has a component that is proportional to the $H_{\text{ext}}^{-1}$, indicating the presence of a sizable $H_{\text{DL}}$. Surprisingly, the $H_{\text{ext}}^{-1}$ component of $A(H_{\text{ext}})$ inverts its sign suggesting that the sign of $H_{\text{DL}}$ depends on the Py thickness on the H-diamond. The underlying physics is discussed later.

Hereafter, $H_{DL}$ and $H_{FL}$ are discussed as the damping-like torque efficiency $\xi_{DL}$, by accounting the $t_{Py}$ dependence of the electrical current in the 2DHG, $I_{2DHG}$. The torque efficiencies of the damping-like and the filed-like toques are described as follows[34], $\xi_{DL} = \left(\frac{2e}{\hbar}\right)\mu_0 M_S t_{Py} t_{2DHG} w \frac{H_{DL}}{I_{2DHG}}$, where $w$ is the width of the channel, $M_S$ is the saturation magnetization, and $t_{2DHG}$ is the thickness of 2DHG. Since the 2DHG in H-diamonds exist within 1 nm of the top surface[20,35], we postulate the thickness of the 2DHG to be 1 nm, which allows the estimation of the minimum value of $\xi_{DL}$. Note that $I_{2DHG}$ is estimated by using a conventional the parallel circuit model. Figure 3(d) represents the whole dataset of the $t_{Py}$ dependence of $\xi_{DL}$ of Py on H- and O-diamonds. Given that the contribution of the self-induced SOT in Py increases with increasing $t_{Py}$ and the sign of SOT of Py is positive[36], the sign of the SOT efficiency of the H-diamond is negative[37]. Meanwhile, negligibly small SOT is observed in O-diamond that is insulative. However, the self-induced SOT of Py on H- and O-diamonds could be different because the self-induced SOT is attributed to the asymmetric spin accumulation at the top and bottom surface of the Py layer. Therefore, a small negative value $H_{DL}$ recorded for the 4 nm-thick Py on O-diamond device can be explained by the self-induced SOT in the Py layer.

To obtain a more complete understanding of the sizable SOT observed for the H-diamond/Py, the $t_{Py}$ dependence of $\xi_{DL}$ is calibrated by using a simple spin-diffusion equation in a 2DHG/ferromagnet bilayer. The model is based on the conventional spin-diffusion model[37,38], and the 2DHG is introduced as a boundary condition of bottom side of the ferromagnet; spin accumulation potential at the bottom is generated by Edelstein effect[39-41], $<\delta\mu_s> = \lambda_{REE}E$, where $\lambda_{REE}$ is charge-spin conversion index for the Rashba-Edelstein effect (the Rashba-Edelstein effect length), and $E$ is the electric field in 2DHG generated by injecting the ac current for the harmonic Hall measurement, respectively[28]. Here, we assumed that the spin relaxation time of 2DHG is shorter or comparable to the momentum relaxation time to introduce the spin current absorption into the 2DHG[41]. By considering the self-induced SOT in the Py layer, the spin current density $J_{sI}$ at the interface of the H-diamond and the Py is described as follows[37]:

$$J_{sI} = \left( \frac{\sigma_{FM}}{\sigma_{2DHG}} \frac{1}{\lambda_{FM}} \tanh\left(\frac{t_{FM}}{\lambda_{FM}}\right) \lambda_{REE} + \frac{\sigma_{FM}}{\sigma_{2DHG}} \left(1 - \frac{1}{\cosh\left(\frac{t_{FM}}{\lambda_{FM}}\right)}\right) \theta_{FM} \right) J_{c(2DHG)}, \qquad (2)$$

where $\lambda_{Py}$ is the spin diffusion length of Py, $\sigma_{Py(2DHG)}$ is the electrical conductivity of Py (2DHG), $J_{c(Py)}$ ($J_{c(2DHG)}$) is the electrical current density in Py (2DHG), and $\theta_{Py}$ is the spin Hall angle of Py. $\xi_{DL}$ is expressed as $\xi_{DL} = J_{sI} / J_{c(2DHG)}$ so that $t_{Py}$ dependence of $\xi_{DL}$ can be expressed by the model considering the self-induced SOT in the Py layer. For simplicity, the spin precession during spin current diffusion and the imaginary part of the spin-mixing conductance are neglected[42,43]. Figure 4 shows the experimental results of the $t_{Py}$ dependence of $\xi_{DL}$ of the H-diamond/Py and the results of the aforementioned model calculation. Here, $\xi_{DL}$ is calculated for $\sigma_{2DHG} = 2.8 \times 10^{-5}$ S/□, corresponding to $\sigma_{2DHG}$ of the bare H-diamond[28]. The $t_{Py}$ dependence of $\sigma_{Py}$ is estimated from the experimental results and fitting[28]. The experimental data was fitted with parameters $\lambda_{REE} = -0.19 \pm 0.07$, $\theta_{Py} = 0.07 \pm 0.02$, and $\lambda_{Py} = 3.78 \pm 1.57$ nm, respectively, and the parameters of Py are comparable to the previous study[36]. The results consort an understanding that giant charge-spin conversion in H-diamond can be rationalized from the $t_{Py}$ dependence of $\xi_{DL}$ in the H-diamond/Py devices. The negative $\lambda_{REE}$ obtained by the fitting is ascribed to the fact that a direction of built-in potential inducing the 2DHG is opposite to that of 2DEG, i.e., the upward band bending occurs unlike the downward band bending in 2DEG. The emergence of the giant spin conversion efficiency in the material system consisting of only light and ubiquitous elements, carbon and hydrogen, can surmount limitation of material selection shedding light mainly on heavy and/or rare elements, which is reminiscent of novel green technology applications of this materials system.

Comparison of the amplitude of charge-to-spin conversion efficiency in H-diamond with other SOT materials can underpin a distinct advantage of this material system. The index of charge-spin conversion efficiency in a Rashba system, the inverse Rashba Edelstein length $\lambda_{IREE}$, was reported to be about 0.2 nm at room temperature in the LAO/STO interface [44,45], which are comparable to the $\lambda_{REE}$ in the H-diamond/Py, which signify that the 2DHG in H-diamond exhibits comparably giant spin-orbit torque on-par with the oxide-based 2DEGs (we note that $\lambda_{REE}$ has been also compared with $\lambda_{eff}$, which is an index of charge spin conversion typically described as the product of the spin diffusion length and spin Hall angle and is constant in a certain material with an intrinsic contribution to the spin Hall effect[27,45]). Meanwhile, of substantial importance is the fundamental difference that electrical transport in 2DHG is mediated by holes, compared to electron transport in the oxide 2DEGs was electron, and more importantly, the fact that our material system allows direct contact to the 2DHG of H-diamond unlike the case of LAO/STO. Notably, the top surface of STO can also host a 2DEG, which may circumvent the problem of difficulty in the direct access[46]. Meanwhile, large atomic numbers, $Z$, of Sr ($Z$=38) results in giant spin-orbit coupling in an atom comparing with the SOC in carbon ($Z$=6) and hydrogen ($Z$=1) according to the approximate $Z^4$-rule as a measure of atomic SOC. The magnitude of the charge-to-spin conversion efficiency has been discussed in view of $\lambda_{eff}$ in a wide variety of materials at RT: semimetal (MoTe$_2$, > 1 nm)[47], topological insulator (Bi$_2$Se$_3$, 10 nm at room temperature)[27,48], and heavy metal (Pt, 0.05 nm)[27]. Here, $\lambda_{REE}$ is estimated to be 0.19 nm at RT in the H-diamond/Py system by the best fit of Fig. 4, which is several times greater than that of heavy metals[28] despite H-diamond consists of elements with inherently small SOI.

In summary, the emergence of giant spin torque using 2DHG at H-diamond was successfully demonstrated at RT, where the strong Rashba SOI in the 2DHG was exerted. Control experiments using insulating O-diamond and the investigation of the thickness dependence in the top Py layer on the H-diamond of $\xi_{DL}$ bear out our assertion. Despite the H-diamond consists of only light elements, the index of the charge-spin conversion efficiency, $\lambda_{REE}$, is several times greater than that in heavy elements, which paves a pathway to construct novel green-technology-based material platforms for spintronics and spin-orbitronics.

This work is supported in part by a Grant-in-Aid for Scientific Research (S) (No. 16H06330), Grant-in-Aid for Scientific Research (A) (No. 21H04561), MEXT Initiative to Establish Next-Generation Novel Integrated Circuits Centers (X-nics, Tohoku Univ., Japan) and MEXT Q-LEAP (No. JPMXS0118067395). H.H. acknowledges financial support by the Deutsche Forschungsgemeinschaft (DFG, German Research Foundation) via Germany's Excellence Strategy EXC-2111- 390814868 and TRR 360 (Project-ID 492547816).

**Figure & Table Captions**

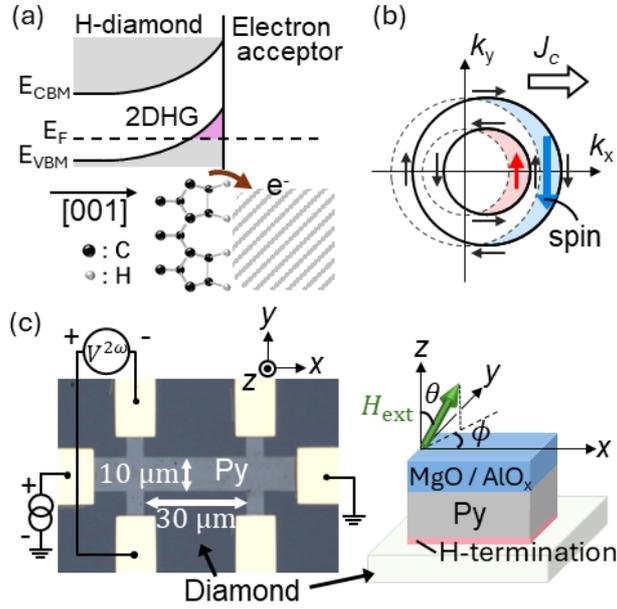

**Fig. 1 (a)** Schematic illustration of the surface of hydrogen-terminated diamond (H-diamond) and its band diagram. Electron acceptors at the surface extract electrons from the H-diamond surface, resulting in the band bending and the 2DHG formation. **(b)** Spin texture and spin splitting appearing in the H-diamond surface due to the Rashba spin-orbit interaction (SOI) and spin accumulation due to the applied electric field to the 2DHG. **(c)** Optical image of the H-diamond/Py device and measurement setup of the second harmonic Hall measurement. The external magnetic field with the zenith angle, $\theta$, and the azimuthal angle, $\phi$ is applied to the Hall bar device.

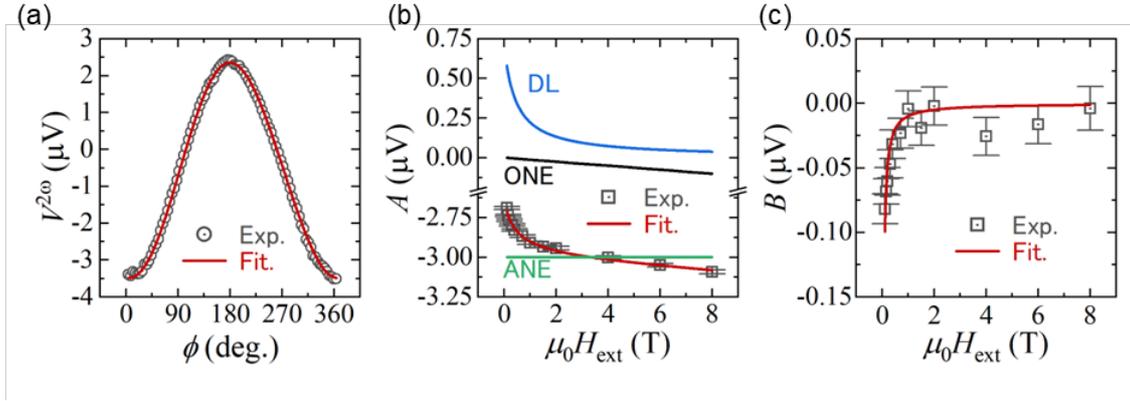

**Fig. 2 (a)** The azimuthal-angle $\phi$ dependence of the second harmonic Hall voltage $V^{2\omega}$ in the H-diamond/Py(4 nm) device. The black open circles are experimental results and the red solid curve is the fitting result with Eq. (1) in the main text. The $\mu_0 H_{\text{ext}}$ dependence of **(b)** $A(H_{\text{ext}})$ and **(c)** $B(H_{\text{ext}})$ components in Eq. (1) of the H-diamond/Py(4 nm) device, where the black open squares show the experimental results and the red solid line is the fitting line. In **(b)**, the damping-like torque (DL), the ordinary Nernst effect (ONE), and the anomalous Nernst effect (ANE) contributions in the result are deconvoluted by the fitting and shown as the blue, black, and green solid lines, respectively.

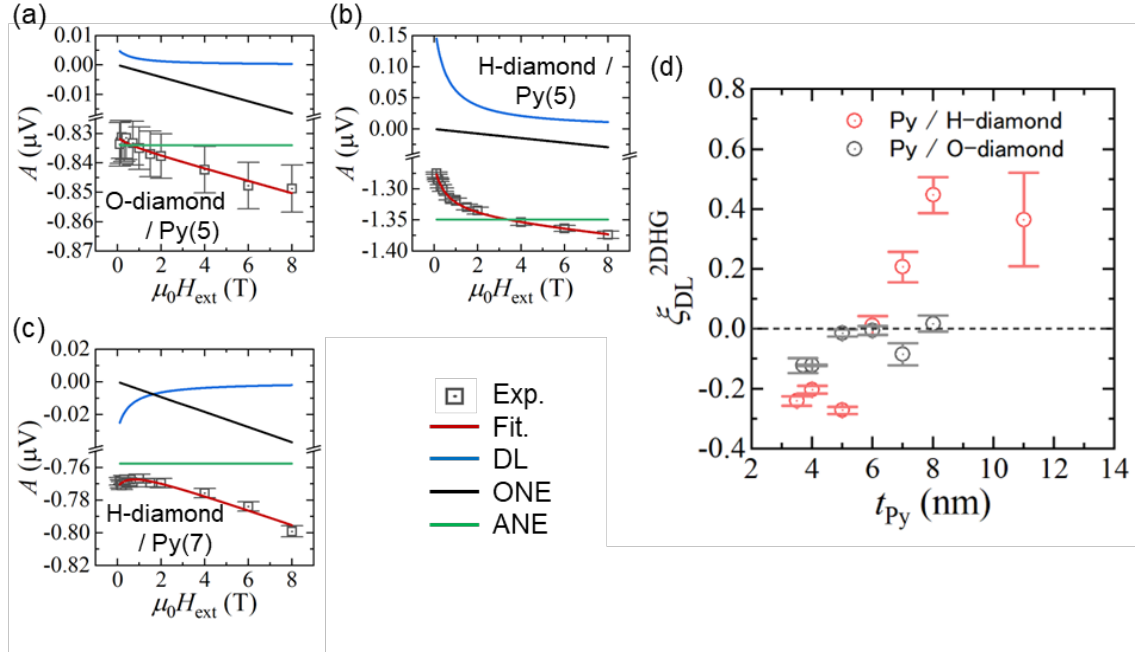

**Fig. 3** The $\mu_0 H_{\text{ext}}$ dependence of $A(H_{\text{ext}})$ of **(a)** O-diamond/Py(5 nm), **(b)** H-diamond/Py(5 nm), and **(c)** H-diamond/Py(7 nm), respectively. The black open squares are experimental results and the red solid curve is the fitting result. The damping-like torque (DL), the ordinary Nernst effect (ONE), and the anomalous Nernst effect (ANE) contributions in the result are deconvoluted by the fitting and shown as the blue, black, and green solid lines, respectively. **(d)** The Py thickness $t_{\text{Py}}$ dependence of $\xi_{\text{DL}}$ in the H-diamond/Py devices. The red (black) open circles show the experimental data measured from the H-diamond (O-diamond) devices. The black dotted line indicates $\xi_{\text{DL}} = 0$.

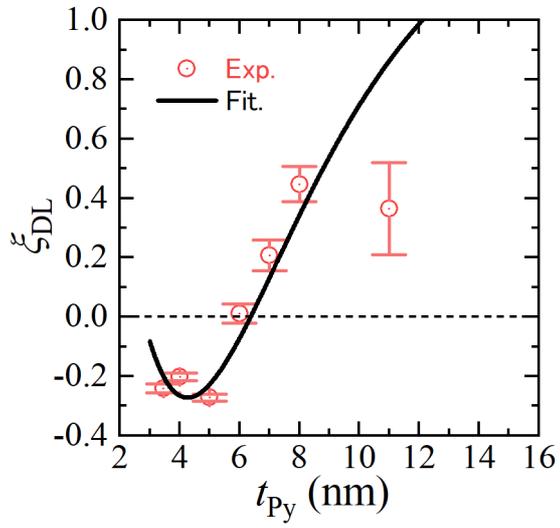

**Fig. 4** The $t_{Py}$ dependence of $\xi_{DL}$ of the H-diamond/Py devices (the red open circles). The black solid line shows the fitting line obtained by using Eq. (2) in the main text. The calculation reproduces the data when $\lambda_{REE} = -0.19 \pm 0.07$ nm, $\theta_{Py} = 0.07 \pm 0.02$, and $\lambda_{Py} = 3.78 \pm 1.57$ nm, respectively. The black dot line indicates $\xi_{DL} = 0$.